\documentclass[aps,prc,preprint,superscriptaddress,floatfix]{revtex4-1}
\usepackage{amsmath}
\usepackage[english]{babel}
\usepackage[bookmarks=false,colorlinks=false]{hyperref}
\usepackage{graphicx}
\usepackage{url}
 \usepackage{footnote}

\begin{document}

\title{On the Spiral Structures in Heavy-Ion Collisions}

\author{A.~Rustamov}
\affiliation{Frankfurt Institute for Advanced Studies\\
Goethe University Frankfurt, D-60438  Frankfurt am Main, Germany}
 \affiliation{National Nuclear Research Center,\\
 AZ-1073 Baku, Azerbaijan}

\author{J. N.~Rustamov}
\affiliation{Shamakhy Astrophysical Observatory,\\ 
AZ-5626 Shamakhy, Azerbaijan}

\begin{abstract}
It is  well established that many galaxies, like our Milky Way,  exhibit spiral patterns. The entire galactic disc rotates about the galactic centre with different speeds;  higher closer to the centre, lower at greater distances - that is, galactic discs do not rotate like a solid compact disc. The spiral arms are the part of the galactic disc where many young stars are being born. Since young stars are also brightest, we can see the spiral structure of other galaxies from afar. Typically spiral galaxies are copiously observed at redshifts $z \approx 1$. The recently observed grand-design galaxy Q2343-BX442 at $z=2.18$, however, implies uncertain origin of its spiral structure~\cite{sg}. Indeed such "old" galaxies usually look rather clumpy  because of their dynamically hot discs. In this report, based on self-similarity, we argue that spiral structures may also appear in heavy-ion collisions as messengers of  phase transitions. Thus, spiral structures in galactic patterns may be traced back to a few microseconds after the Big Bang.  

\end{abstract}

\maketitle

\section{Introduction}

To our knowledge Rolf Hagedorn was the first to consider the concept of  self-similarity in particle physics, which ultimately led to the concept of limiting temperature and phase transitions in strongly
interacting matter~\cite{Hagedorn1, Hagedorn2, Hagedorn3, Hagedorn4}.  His main idea was to search for self-similar patterns in the composition of heavy particles; heavy particles are composed of lighter ones, and these again in turn of still lighter one, until one reaches a fundamental constituent, which in his case was  a pion, the lightest hadron~\cite{redlich}. Mathematically the problem is formulated with the following bootstrap equation \cite{Fraurschi}:

\begin{equation}\label{SB}
        \rho(m,V_{0})  =  \delta(m - m_{0}) + \sum_{N=2}^{\infty} \frac{1}{N!}\left[\frac{V_{0}}{(2\pi)^3}\right]^{N-1}\prod_{i=1}^{N}\int dm_{i}\rho(m_{i})\int d^3p_{i}\delta^{4}\left(\sum_{i}p_{i}-p\right),
         \end{equation}
here the first delta function represents pion, while the last  four-delta function guaranties the 4-momentum conservation. The solution of the bootstrap equation is given in Ref.~\cite{Nahm} :

   \begin{equation}
        \rho(m,V_{0})  \sim m^{-3}e^{m/T_{H}},
         \end{equation}
where $T_{H}$ is referred to as Hagedorn limiting temperature. By taking the composition volume $V_{0}$ to be spherical with the radius of inverse pion mass Hagedorn got a value of about 150 $MeV$ for the limiting temperature  $T_{H}$.

The self-similar properties of the Hagedorn fireballs within the statistical-bootstrap model
were formulated by Frautschi~\cite{Fraurschi} who used Eq.~(\ref{SB}) to describe the fireball decay. 
For a heavy fireball, $m\gg m_0$, the probability distribution ${\cal P}(n)$ of its one-step decay into $n$
lighter fireballs ($n\ge 2$) appears to be independent of $m$ and has the form:

\begin{equation}
{\cal P}(n)~=~\frac{(\ln 2)^{n-1}}{(n-1)!}~.
\label{dec}
\end{equation}
This equation shows that  dominant channels correspond to 
2-body decays with ${\cal P}(2)\cong 0.69$ and 3-body decays with ${\cal P}(3)\cong 0.24$.
If  $m_1$ and $m_2$ which are the decay products in a reaction $m\rightarrow m_1+m_2$ 
are also heavy, Eq.~(\ref{dec}) applies in this case as well.
Original fireball with mass $m$ and its constituents $m_1$ and $m_2$
have the same properties provided that all these masses are much larger
than $m_0$ and $T_H$. Therefore, one finds a similar behaviour for the whole system $m$ 
and for its part $m_1$, that is what self-similarity stands for.

\section {Most popular self-similar objects}	
Any object  is said to be self-similar if it is reproduced by magnifying some part of it.  Self-similarity can manifest itself both in discrete and continuous fashion, albeit in most cases the exact self similarity is only asymptotically so. A well known example of discrete self-similarity are Russian dolls  where a larger doll discretely hides a similar smaller one inside it and so forth. If we had infinite number of dolls, both ever smaller and ever larger, we would have a set with exact discrete self-similarity. The obvious limitation on the size of the dolls from both sides makes this discrete self-similarity only approximate.  Another charming and copiously encountered in nature example of a self-similar object is the logarithmic  spiral. In polar coordinates the equation of the logarithmic spiral is given as:
	\begin{equation}
         r(\phi) = r_{0}e^{k\phi},
         \label{log_spiral}
         \end{equation}
where $r_{0}$ and $k$ are constants. Scaling the spiral by a factor $s$ is equivalent to the same spiral rotated by a constant angle of $ln(s)/k$:

         \begin{equation}
         sr(\phi) = sr_{0}e^{\left( k\phi \right)} = r_{0}e^{ln(s)}e^{k\phi} = r_{0}e^{k\left( \phi + \frac{ln(s) }{k} \right)}.
         \end{equation}
For a particular value of $s=e^{2\pi km}$, with $m$ being any  integer number, one gets the same spiral rotated by an angle:

\begin{equation}
        ln(e^{2\pi km})/k = 2\pi m
         \end{equation}

As in the case of Russian dolls, the exact self-similarity for the logarithmic spiral is limited by finite value of $r_{0}$ (cf. Eg.~\ref{log_spiral}) and the size of the spiral. 
\section {Spiral structures  in strongly interacting matter}
In non-central high energy heavy-ion collisions large orbital angular momenta are developed, with its vector directed perpendicularly
to the reaction plane. The latter is defined by the beam direction and the impact parameter. A fraction of the angular momenta is
taken away by the spectators, though the strongly interacting matter in the overlap region will also carry a substantial orbital angular momenta.
For instance, in a non-central Au+Au collisions at RHIC energies the global angular momentum of the overlapping matter was estimated to be of the
order of $10^5$ spin units~\cite{orbmom}. Consequently the produced matter will rotate.
This should be manifested in the polarisation of the produced spin non-zero hadrons.
In particular, polarisation of vector mesons and hyperons are expected. On the other hand,
a correlation length diverges near the critical point, and the matter has to rotate only in a scale invariant, i.e self-similar way.
This scenario can be achieved if  rotations follow a logarithmic spiral mentioned in the previous section.
Experimentally such structures can probably be looked at in event-by-event  angular and momentum distributions of measured particles. We plan to address possible signatures of spirals in heavy ion collisions in following papers.
\section {Conclusion}
In summary, near the critical point, spiral structures should appear in the non-central heavy-ion collisions. This scenario may arise when combining two distinct arguments 
on non-vanishing orbital angular momentum in heavy-ion collisions and well known phenomenon of the  divergence of the correlation length near the critical point. 
While the first argument leads to the overall rotation of the system in the reaction plane,
the second one requires the scale invariance, i.e., self-similarity of the system. The logarithmic spiral satisfies both of these requirements. 
On the other hand, the constant shape of the logarithmic spiral reveals itself in nature at all scales; from unicellular foraminifera, nautilus shell, sunflower seeds up to giant structures in the Universe. One can argue that these structures may be artefacts of successive phase transitions the Universe passed through during its first moments.
\section{ACKNOWLEDGMENTS}
A. R. would like to thank Marek Ga\'zdzicki and Mark Gorenstein for critical discussions and comments.

\end{document}